\documentclass{iopjournal}
\usepackage{lmodern}
\usepackage{amssymb}
\usepackage[nosort,compress]{cite}

\newcommand{\mydot}[1]{\rlap{\raisebox{1.4ex}{\hspace{0.2em}$\cdot$}}#1}

\newcommand{\myddot}[1]{\rlap{\raisebox{1.4ex}{\hbox to 0.4em{\hfil$\cdot$\hfil$\cdot$\hfil}}}#1}

\newcommand{\mydddot}[1]{\rlap{\raisebox{1.4ex}{\hbox to 0.4em{\hfil$\cdot$\hfil$\cdot$\hfil$\cdot$\hfil}}}#1}

\begin{document}

\articletype{Paper} 

\title{Jerky chiral active particles}

\author{Stephy Jose$^{1,*}$\orcid{0000-0002-5089-9873} and Hartmut L{\"o}wen$^{1}$\orcid{0000-0001-5376-8062}}

\affil{$^1$Institut f{\"u}r Theoretische Physik II: Weiche Materie,
Heinrich-Heine-Universit{\"a}t D{\"u}sseldorf, D-40225 D{\"u}sseldorf, Germany}

\affil{$^*$Author to whom any correspondence should be addressed.}

\email{stephy.jose@hhu.de, hlowen@thphy.uni-duesseldorf.de}

\keywords{jerk, chirality, active matter, active Brownian particle, circle swimmers}

\begin{abstract}
We introduce jerky chiral active Brownian particles (jcABPs), a generalization of conventional chiral active Brownian particles (cABPs) subjected to jerk, the time derivative of acceleration, and analytically derive their mean displacement and mean squared displacement (MSD). Our results show that jerk induces anomalous fluctuations and oscillatory behavior on the standard circular swimming of chiral active particles. The interplay of jerk, chirality and persistence produces a family of mean trajectories including damped and exploding Lissajous patterns alongside the well-known \emph{spira mirabilis} (logarithmic spirals). Our work on jerky chiral active particles opens a new route to explore rich dynamical effects in active matter.
\end{abstract}

In many physical systems, memory can arise from coarse-graining over hidden degrees of freedom, delayed feedback, history-dependent friction, or time-correlated noise~\cite{ Mori_1965, Zwanzig_1961,Mori_1965b, Zwanzig_1970, Espanol_1993, Bocquet_1994, Mason_1995,Nakajima_1958,leung2014bifurcation,keim2019memory,Kowalik_2019, Straube_2020, goychuk2022memory,Scalfi_2023,Mitterwallner_2020, klimek2022optimal,klimek2024data,dalton2025memory,milster2025nonequilibrium}. One minimal way to model such memory is to include the time derivative of acceleration, or \emph{jerk}, in the equations of motion~\cite{schot1978jerk,linz1997nonlinear,eichhorn1998transformations,eichhorn2002simple,umut2013simple,rajagopal2018autonomous,Wu2006,Gottlieb1998,Gottlieb2004,Gottlieb2006}. For instance, in colloidal systems immersed in viscoelastic or non-Newtonian fluids, the frictional force is not instantaneous but instead depends on the entire history of particle through a memory kernel~\cite{GomezSolano2015,Puertas2014,Narinder2018,Treffenstaedt2020,Dalton2024,Sprenger2022}. Expanding such memory kernels leads to terms proportional to jerk~\cite{lowen2025gigantic}. Advances in engineering have enabled the experimental realization of feedback-controlled systems using synthetic forces leading to effective dynamics with time delays~\cite{bechhoefer2005feedback,van2022real,saha2024cybloids,holubec2022equilibrium,kopp2023spontaneous
}. When an external force is applied based on delayed measurements of the motion of a particle, the resulting dynamics can involve an explicit dependence on the acceleration of the particle and its rate of change~\cite{lowen2025gigantic}. Such systems challenge the foundational assumptions of classical mechanics, such as the force depending only on the velocity or the positivity of the inertial mass~\cite{newton1833philosophiae}. They describe physically realizable systems with rich and often counterintuitive dynamics.

Far from equilibrium, memory becomes even more prominent~\cite{loos2025non,paulsen2019minimal}. The presence of memory fundamentally alters transport, fluctuations, and response in stochastic systems~\cite{zhao2024emergence,dieball2022scattering}. It can induce anomalous diffusion, modify relaxation spectra, and lead to long-range temporal correlations~\cite{yulmetyev2000stochastic,kim2014anomalous,stanislavsky2015anomalous,korsakas2020long,bhattacharyya2025nonequilibrium}. A paradigmatic example is active matter~\cite{marchettireview,ramaswamyreview,elgetireview,Bechinger2016b,Gompper2020} composed of self-driven entities that consume energy to generate motion. Examples of active matter span a wide range of scales and materials from natural systems such as motile bacteria and eukaryotic cells~\cite{bergbook,tailleur2008}, to artificial systems such as synthetic colloidal swimmers~\cite{ebbens2010active,golestanian2005propulsion,howse2007self} and active granular rods~\cite{kudrolli2008swarming,deseigne2010collective}. A central goal in active matter physics is to understand how simple rules of individual propulsion and interaction give rise to complex collective dynamics and non-equilibrium phases. A widely studied model is the \emph{active Ornstein–Uhlenbeck process} (AOUP), where self-propulsion arises from temporally correlated (colored) noise, endowing the particle with a finite persistence time, a form of memory in its motion~\cite{Uhlenbeck1930,Szamel2014,Bonilla2019,Martin2021,Dabelow2021,Wittmann2018,Fritz2023,Gupta2023,Crisanti2023}. Another simple microscopic model of active matter is the overdamped \emph{active Brownian particle} (ABP), in which a particle moves with a fixed propulsion speed along a fixed direction that undergoes rotational diffusion~\cite{fily2012athermal,lindner2008diffusion,basu2018active,romanczuk2010collective,romanczuk2012active}. Despite its simplicity, these models capture a wide range of interesting phenomena such as enhanced diffusion, ballistic-diffusive crossover, boundary accumulation, and motility induced phase separation~\cite{elgetireview,cates2013active,merrigan2020arrested,lee2013active,jose2023current}. 

More recently, significant attention has turned to models incorporating \emph{inertia} and underdamped ABPs have been proposed to describe active particles in dilute gases, vacuum, or low-viscosity fluids~\cite{scholz2018inertial,sprenger2023inertial,caprini2021active,caprini2022rotational,sprenger2023dynamics,sprenger2021rocket,lowen2020inertial,gutierrez2020inertial,nguyen2021active,montana2023inertial,caprini2021inertial}. They have a mass and characterize dynamics over both the inertial timescale and the rotational persistence timescale. Inertia leads to persistent velocity correlations~\cite{caprini2021spatial,marconi2021hydrodynamics}, transient superdiffusion~\cite{nguyen2021active,sprenger2021rocket}, and effective temperatures~\cite{hecht2024define,caprini2023entropons}. In parallel, active matter models with \emph{chirality} have been introduced to capture the effects of broken rotational symmetry on the dynamics of self-propelled particles\cite{teeffelen2008dynamics,deseigne2010collective,Gompper2020,sprenger2021rocket,Nosenko_2020,zhang2020reconfigurable,baconnier2025self,bayati2022memory,pattanayak2024impact}. These models typically prescribe a constant angular velocity to the orientation dynamics, resulting in circular trajectories in the absence of rotational noise~\cite{teeffelen2008dynamics,kummel2013circular,jahanshahi2017brownian}. The noise-averaged mean displacement follows a \emph{spira mirabilis} or logarithmic spiral~\cite{teeffelen2008dynamics,sprenger2021rocket}. Chiral active particles display a range of interesting behaviors, including spiral drift, circular confinement, and complex scattering dynamics~\cite{teeffelen2008dynamics}. Beyond inertia and chirality, a new class of active particles incorporating jerk, the time derivative of acceleration has recently been introduced~\cite{lowen2025gigantic}. This is an extension of an active Ornstein-Uhlenbeck model and  displays anomalously large fluctuations and super-ballistic scaling of the mean-squared displacement.

In this work, we introduce a \emph{jerky chiral} ABP (jcABP) model by incorporating jerk in the translational equation of motion and chirality via a constant angular velocity in the orientational equation of motion in two dimensions. This model enables us to systematically investigate the impact of jerk, inertia, and chirality in active particle dynamics. We analytically derive the Green's function, and exact expressions for mean displacement and MSD in various limiting cases. We demonstrate that jerk induces transient oscillations in both the mean trajectory and the mean-squared displacement, even in the absence of chirality. When combined with chirality, jerk fundamentally alters the circular swimming behavior of active particles. Specifically, we find that the radius of the emergent orbit becomes a non-trivial function of the jerk time scale. In the presence of rotational noise, the interplay between jerk, chirality, and persistence leads to spiral trajectories resembling \emph{Lissajous} curves. The shape of the trajectories crucially depend on the relative values of different underlying parameters. We also show that jerk introduces anomalous fluctuations characterized by unusually high power-law exponents at short times.

This paper is organized as follows. In Section~\ref{sec:model}, we introduce the model. In Section~\ref{sec:greens_fn}, we derive the orientation correlation functions and the Green’s function. In Section~\ref{sec:vanishing_chirality}, we consider the limiting case of vanishing chirality and obtain exact expressions for the mean displacement and the MSD. Here we show that jerk induces oscillatory behavior in the cumulants of the displacement and anomalous short-time fluctuations characterized by a $t^5$ scaling. In Section~\ref{sec:finite_chirality}, we analyze the effects of finite chirality, treating separately the cases with and without rotational noise. Without rotational noise, the particle follows circular trajectories at long times, with a radius that depends non-trivially on chirality as well as the friction and jerk timescales. In the presence of rotational noise, we obtain spiral trajectories which resemble intricate Lissajous patterns with beats and modulations. Depending on the parameter values, these curves can be either damped or exploding. We present conclusions and discussion in Section~\ref{sec:conclusion}. Finally, details pertaining to a few calculations in the main text are provided in Appendices~\ref{app:app1}-\ref{app:app2}.

\section{Model}
\label{sec:model}

We consider a two-dimensional active Brownian particle subject to jerk, inertia, and chirality, driven by both self-propulsion and thermal noise. The position of the particle $\vec{r}(t) = (x(t), y(t))$ evolves according to the third-order Langevin equation:
\begin{equation}
\lambda  \mydddot{\vec{r}}(t) + m \myddot{\vec{r}}(t) + \gamma \mydot{\vec{r}}(t)
= \gamma v_0 \hat{n}(t) + \sqrt{2D \gamma^2}  \vec{\eta}(t),
\label{eq:jerky_lang}
\end{equation}
where $\lambda$ is the coefficient of the jerk, $m$ is the mass of the particle, and $\gamma$ is the viscous friction coefficient. The right-hand side includes a self-propulsion force of magnitude $\gamma v_0$ along the instantaneous orientation $\hat{n}(t) = (\cos \theta(t), \sin \theta(t))$, as well as a stochastic thermal force modeled by a Gaussian white noise $\vec{\eta}(t)$ with vanishing mean $\langle \eta_i(t) \rangle = 0$ and delta correlations $\langle \eta_i(t') \eta_j(t'') \rangle = \delta_{ij} \delta(t' - t'')$. Also, $v_0$ is the constant self-propulsion velocity and $D$ is the translational diffusion constant. 

The orientation angle $\theta(t)$ evolves independently via angular velocity $\omega_0$ (chirality) and a stochastic rotational diffusion term:
\begin{equation}
\mydot{\theta}(t) = \omega_0 + \sqrt{2 D_r} \, \eta_r(t),
\label{eq:theta_dyn}
\end{equation}
where $D_r$ is the rotational diffusion constant and $\eta_r(t)$ is a Gaussian white noise with vanishing mean $\langle \eta_r \rangle=0$ and delta correlation $\langle \eta_r(t') \eta_r(t'') \rangle = \delta(t' - t'')$. The angular noise drives decorrelation of the orientation over a characteristic time scale $1/D_r$, while $\omega_0$ introduces persistent circular motion and chirality. In this framework, the particle’s position $\vec{r}(t)$ describes a stochastic trajectory in the $x$--$y$ plane. The instantaneous magnitude of the position vector is given by $r(t) = |\vec{r}(t)| = \sqrt{x(t)^2 + y(t)^2}$, and the azimuthal angle in real space is $\phi(t) = \tan^{-1}(y(t)/x(t))$. Note that $\phi(t)$ is distinct from the internal propulsion angle $\theta(t)$, which governs the direction of self-propulsion.

We identify the relevant time scales in the system: the jerk time scale $
\tau_J = {\lambda}/{m}$, the friction time scale $\tau_F = \sqrt{{\lambda}/{\gamma}}
$, the persistence time scale of the rotational noise $\tau_P=1/D_r$, and the chiral time scale $\tau_C=1/\omega_0$. In terms of these quantities, equation~(\ref{eq:jerky_lang}) becomes
\begin{equation}
 \mydddot{\vec{r}}(t) + \myddot{\vec{r}}(t)/\tau_J + \mydot{\vec{r}}(t)/\tau_F^2
=(\gamma v_0/\lambda) \, \hat{n}(t) + \sqrt{2 D \gamma^2/\lambda^2} \, \vec{\eta}(t).
\label{eq:lang_time_scales}
\end{equation}
We take the persistence time $\tau_P$ as the natural time unit and the persistence length $l_P=v_0 \tau_P$ as the natural length unit.
For simplicity, we assume that the particle is completely at rest at time $t=0$ and consider initial conditions of the form:
\begin{equation}
\vec{r}(0) = \mydot{\vec{r}}(0) = \myddot{\vec{r}}(0) = \vec{0}~\mathrm{and}~\theta(0)=0.
\label{init_conds}
\end{equation}
We next systematically analyze equation~(\ref{eq:lang_time_scales}) to derive the Green's function, and cumulants of the displacement $\vec{r}(t)$.


\section{Orientation correlations and Green's function}
\label{sec:greens_fn}
The orientation correlation function is given as~\cite{sprenger2021rocket,scholz2018inertial}~(see Appendix~\ref{app:app1} for a derivation)
\begin{equation}
\langle \hat{n}(t') \cdot \hat{n}(t'') \rangle = \cos[ (t' - t'')/\tau_C]e^{-|t' - t''|/\tau_P} .
\label{orientation_correlation}
\end{equation}
This correlation function captures both the exponential decay due to rotational diffusion and the oscillatory behavior due to chirality. As we will see, it enters as a kernel in the evaluation of the cumulants of the displacement. Let us define the Green's function ${G}(t)$ as the response to a $\delta$-kick. Due to isotropy and linearity, the Green's function is diagonal and scalar-valued:
\begin{equation}
    \lambda \mydddot{G}(t) + m \myddot{G}(t) + \gamma \mydot{G}(t) = \delta(t).
    \label{eqn:greens_fn}
\end{equation}
To proceed, we define the one-dimensional Fourier transform of a function $f(t)$ and its inverse $\tilde{f}(\omega)$ as
\begin{eqnarray}
\tilde{f}(\omega) &=& \nonumber\int_{-\infty}^{\infty} dt \, e^{i \omega t} f(t), \\
f(t) &=& \frac{1}{2\pi} \int_{-\infty}^{\infty} d\omega \, e^{-i \omega t} \tilde{f}(\omega).
\end{eqnarray}
Taking the Fourier transform of both sides of equation~(\ref{eqn:greens_fn}) yields
\begin{equation}
(i \omega)^3 \lambda \tilde{G}(\omega) + (i \omega)^2 m \tilde{G}(\omega) + (i \omega) \gamma \tilde{G}(\omega) = 1.
\end{equation}
So the Green's function in Fourier space is
\begin{equation}
\tilde{G}(\omega) = \frac{1}{i \lambda \omega^3 - m \omega^2 - i \gamma \omega}.
\label{eq:green_fourier}
\end{equation}
The time-domain response is obtained by computing the inverse Fourier transform:
\begin{equation}
G(t) = \frac{1}{2\pi} \int_{-\infty}^{\infty} 
\frac{e^{-i \omega t}}{i \lambda \omega^3 - m \omega^2 - i \gamma \omega} \, d\omega.
\label{eq:greens_fun_inverse}
\end{equation}
We evaluate this integral by contour integration in the complex $\omega$--plane.
The denominator has poles at $\omega=0$ and at the two nonzero roots of the characteristic equation,
\begin{equation}
i \lambda \omega^3 - m \omega^2 - i \gamma \omega = 0,
\label{characteristic_eq}
\end{equation}
which are
\begin{equation}
\omega_{1,2} = -\frac{i}{2\tau_J} \pm 
\sqrt{\frac{1}{\tau_F^2} - \frac{1}{4\tau_J^2}} \, .
\label{omega_j}
\end{equation}
Applying the residue theorem,
the solution of the third-order linear equation for the Green’s function takes the form
\begin{equation}
G(t) = \sum_{j=1}^{3} C_j e^{-i \omega_j t}.
\end{equation}
The prefactors $C_j$ are determined by matching the continuity conditions of $G(t)$ and its derivatives at $t=0$. Keeping only the causal ($t>0$) contributions,
we obtain 
\begin{equation}
G(t) = \frac{\Theta(t)}{\lambda(\omega_2 - \omega_1)}
\sum_{j=1}^2 (-1)^j \, \frac{1 - e^{-i \omega_j t}}{\omega_j}.
\label{eq:greens_fn_exp}
\end{equation}
which is the explicit expression for the Green’s function for the jcABP dynamics. The nature of the solutions is determined by the real and imaginary parts of $\omega_j~(j=1,2)$ given in equation~(\ref{omega_j}). Writing $\omega_{j}=\omega_R\pm i\,\omega_I$,
the activity-free dynamics is \emph{stable} iff $\omega_I<0$.
When $\tau_F < 2\tau_J$, the square root in equation~(\ref{omega_j}) is real and both frequencies have the same negative imaginary part, $\omega_I= -1/(2\tau_J) < 0$, leading to bounded oscillatory solutions. 
At the critical value $\tau_F = 2\tau_J$, the two nonzero roots become equal and purely imaginary with $\omega_I = -1/(2\tau_J) < 0$, producing non-oscillatory (purely exponential) decay. 
For $\tau_F > 2\tau_J$, the square root becomes purely imaginary: writing $\sqrt{{1}/{\tau_F^2}-{1}/{(4\tau_J^2)}}=i\beta$ with $0<\beta<{1}/{(2\tau_J)}$ gives $\omega_I=-{1}/{(2\tau_J)}\pm\beta<0$, so both modes correspond to distinct decaying exponentials. 
Therefore, when $\lambda,\gamma,m>0$ (so $\tau_J,\tau_F>0$), $\omega_I$ is never positive and the response remains bounded. 
However, exponential growth can occur if the parameter signs are not all positive; for example, a system with negative jerk coefficient, negative inertia, or negative friction admits unstable modes. 
Such cases, while theoretically as practically possible in engineered systems~\cite{lowen2025gigantic}, will be discussed in detail in Section~\ref{Dr_w}.

\begin{figure}[t!]
\centering
\includegraphics[width=1.0\linewidth]{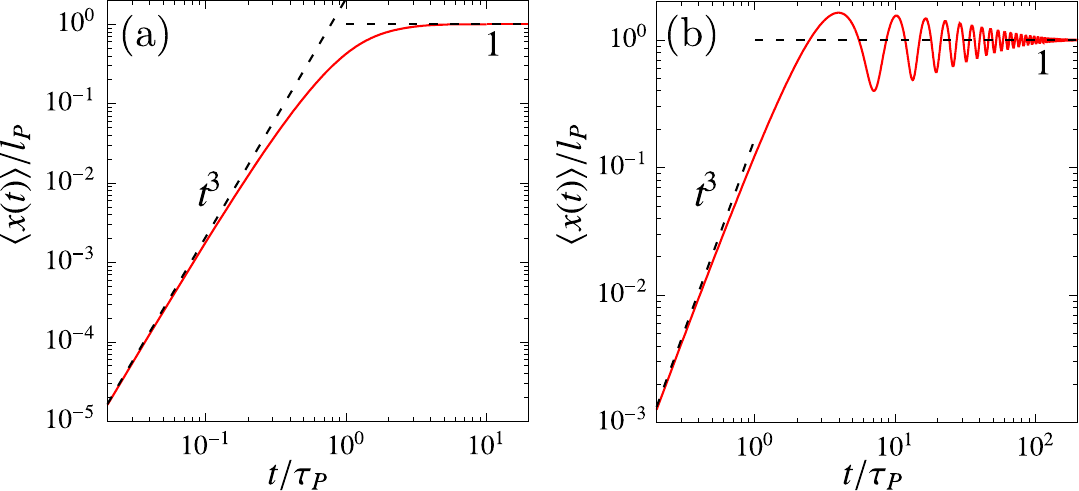}
\caption{Mean displacement $\langle x(t) \rangle$ in units of the persistence length $l_P$ as a function of the reduced time $t/\tau_P$ of a jerky active particle without chirality. The mean displacement grows as $t^3$ at short times and saturates to 
$l_P$ for $t\to\infty$, in agreement with the limits of our exact result provided in equation~(\ref{mean_wo_chirality}).   (a) The parameter values used are $\tau_J=0.2 \tau_P$,~and~$\tau_F=0.28 \tau_P$ corresponding to $\alpha \approx 2.55/\tau_P$. Here, $\tau_P>2\tau_J$ and there is a smooth exponential relaxation of the mean at large times. (b) The parameter values used are $\tau_J=20 \tau_P$, and $\tau_F= \tau_P$ corresponding to $\alpha \approx 1/\tau_P$. Here, $\tau_P<2\tau_J$ and we observe oscillations.
}
\label{fig: mean_without_chirality}
\end{figure}

The  solution of equation~(\ref{eq:jerky_lang}) can be written using the convolution:
\begin{equation}
    \vec{r}(t) = \int_0^t dt' \; G(t - t') \left(\gamma v_0 \hat{n}(t') + \sqrt{2D \gamma^2} \, \vec{\eta}(t')\right).
\end{equation}
Taking an ensemble average, we obtain the mean trajectory:
\begin{eqnarray}
    \langle \vec{r}(t) \rangle &=& \nonumber\gamma v_0 \int_0^t dt' \; G(t - t') \, \langle \hat{n}(t') \rangle \\&=& \gamma v_0 \int_0^t dt' \; G(t - t')e^{- t'/\tau_P}
\left(\begin{array}{c}
\cos( t'/\tau_C) \\
\sin(t'/\tau_C)
\end{array}\right).
\label{eq:MD}
\end{eqnarray}
The mean-squared displacement is then given as
\begin{eqnarray}
\mathrm{MSD}(t) 
  &=&  \nonumber \langle\,|\vec r(t)-\vec r(0)|^2\rangle \\
  &=& \int_0^t dt'\int_0^t dt''\;G(t-t')\,G(t-t'')\,
      \Bigl[\gamma^2 v_0^2\,\langle\hat n(t')\!\cdot\!\hat n(t'')\rangle
            +2D\gamma^2\,\langle\vec\eta(t')\!\cdot\!\vec\eta(t'')\rangle\Bigr]
      \nonumber\\
  &=& 2\gamma^2 v_0^2 
      \int_0^t dt'\!\int_0^{t'} dt''\;\cos\!\bigl[(t'-t'')/\tau_C\bigr]\,
      e^{-(t'-t'')/\tau_P}\,G(t')\,G(t'') 
      \nonumber\\
  && +4D\gamma^2\int_0^tdt' \; G^2(t').
\label{eq:msd}
\end{eqnarray}
The first term accounts for activity-induced fluctuations, while the second term arises from translational Brownian noise. 


\begin{figure}[t!]
\centering
\includegraphics[width=1.0\linewidth]{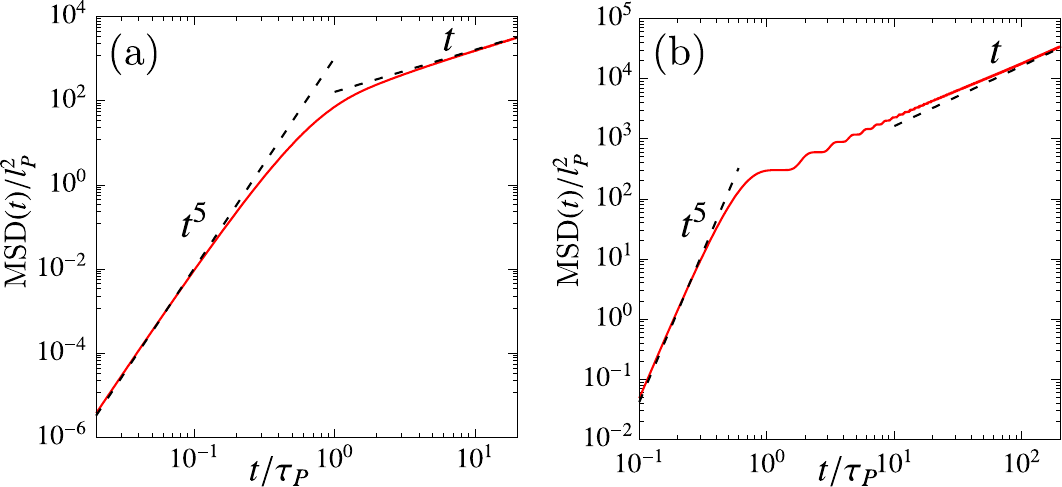}
\caption{Mean squared displacement in units of $l_P^2$ as a function of the reduced time $t/\tau_P$ of a jerky active particle without chirality given in equation~(\ref{msd_zero_omega}). The value $D=0.5$ is held fixed for both plots. The MSD crosses from $\mathrm{MSD}(t)\propto t^5$ at very short times to $\mathrm{MSD}(t)\propto t$ at large times. (a) The parameter values used are $\tau_J=0.2 \tau_P$, and $\tau_F=0.28 \tau_P$ corresponding to $\alpha \approx 2.55/\tau_P$. There is a smooth exponential relaxation of the MSD at large times. (b) The parameter values used are $\tau_J=20 \tau_P$,~and~$\tau_F=0.2 \tau_P$ corresponding to $\alpha \approx 1/\tau_P$. We observe oscillations superimposed on exponential relaxation.
}
\label{fig: msd_without_chirality}
\end{figure}
\section{Special case of vanishing chirality}
\label{sec:vanishing_chirality}

Let us first analyze the case where there is no chirality (i.e., $\omega_0 \rightarrow 0$ or $\tau_C \rightarrow \infty$). This yields the mean displacement
\begin{eqnarray}
    \langle \vec{r}(t) \rangle &= \gamma v_0 \int_0^t dt' \;G(t - t')
\left(\begin{array}{c}
e^{- t'/\tau_P} \\
0
\end{array}\right).
\end{eqnarray}
This asymmetry in $x$ and $y$ directions arises because the initial orientation is aligned with the $x$-axis (i.e., $\theta(0) = 0$), and over time, angular diffusion causes the orientation to decorrelate from this direction. However, since there is no chirality and no initial bias in the $y$-direction, the average motion remains symmetric in $y$, leading to
$
\langle y(t) \rangle = 0.
$
In contrast, the $x$-component retains a net drift due to the initial alignment, resulting in a nonzero mean:
$
\langle x(t) \rangle \neq 0 
$
given as
\begin{eqnarray}
\langle x(t) \rangle 
=\frac{ v_0  \tau_P}{\tau_F^2 \sum_{j=1}^2(-1)^{j}\omega_j} 
\Big[
(1 - e^{-t/\tau_P} )
\sum_{j=1}^2 (-1)^j \frac{1}{\omega_j}
+
\sum_{j=1}^2 (-1)^j \frac{e^{- t/\tau_P} - e^{-i \omega_j t}}{\omega_j(1  - i \omega_j \tau_P)}
\Big].
\label{mean_zero_omega}
\end{eqnarray}
Substituting the values of $\omega_j$ from equation~(\ref{omega_j}), the above expression can be rewritten in the form
\begin{equation}
\langle x(t) \rangle 
= 
{v_0 }{\tau_P }\left[1- \mathcal{A}_1 e^{-t/\tau_P}+ +e^{-t/(2\tau_J)}\left[ \mathcal{B}_1\,\cos(\alpha t)+ \mathcal{C}_1\,\sin(\alpha t) \right]\right]
\;,
\label{mean_wo_chirality}
\end{equation}
where ${v_0 }\tau_P=l_P$ is the persistence length of an active particle and $\alpha$ is the difference in the eigenfrequencies,
\begin{equation}
\alpha=(\omega_1-\omega_2)/2=\sqrt{{1}/{\tau_F^2} - {1}/{(4\tau_J^2)}}.
\label{alpha}
\end{equation}
the prefactors $\mathcal{A}_1,~\mathcal{B}_1$, and $\mathcal{C}_1$ are functions of the underlying timescales $\tau_P,~\tau_J$ and $\tau_F$. We have given the explicit expressions of these prefactors in Appendix~\ref{app:app2}.  
Equation~(\ref{mean_wo_chirality}) has the limiting behaviors
\begin{eqnarray}
\langle x(t) \rangle 
&\buildrel t\to0\over=& 
 \frac{v_0 t^3}{6 \tau_F^2} - \frac{v_0 (\frac{1}{\tau_J} + \frac{1}{\tau_P}) t^4}{24 \tau_F^2}+ \dots~,\nonumber\\
\langle x(t) \rangle 
&\buildrel t\to\infty\over=& 
{v_0 }{\tau_P }=l_P.
\end{eqnarray}
Figure~\ref{fig: mean_without_chirality} shows numerical plots of our analytical expression for the mean displacement given in equation~(\ref{mean_wo_chirality}). The short-time behavior exhibits a super-ballistic regime with $\langle x(t) \rangle \sim t^3$, while at long time, the mean displacement saturates to the persistence length. 
Importantly, the qualitative nature of the dynamics is governed by the values of different time scales. 
If $\alpha$ is imaginary, i.e.,
$
{1}/{\tau_F^{2}} < {1}/{4\tau_J^{2}},
$
the oscillatory terms reduce to purely exponential decay and the relaxation is smooth. 
If $\alpha$ is real, i.e.,
$
{1}/{\tau_F^{2}} >{1}/{4\tau_J^{2}},
$
the system exhibits oscillations that decay over the timescale $2\tau_J$. 
The relative magnitudes of $\tau_P$ and $\tau_J$ determine which decay dominates at long times: 
for $\tau_P > 2\tau_J$, the jerk-induced oscillations decay faster and the final relaxation is smooth on the timescale $\tau_P$~(see figure~\ref{fig: mean_without_chirality}(a)), 
whereas for $\tau_P < 2\tau_J$, the oscillatory envelope on the timescale $2\tau_J$ dominates the long-time behavior~(see figure~\ref{fig: mean_without_chirality}(b)). 
\begin{figure}[t!]
\centering
\includegraphics[width=0.9\linewidth]{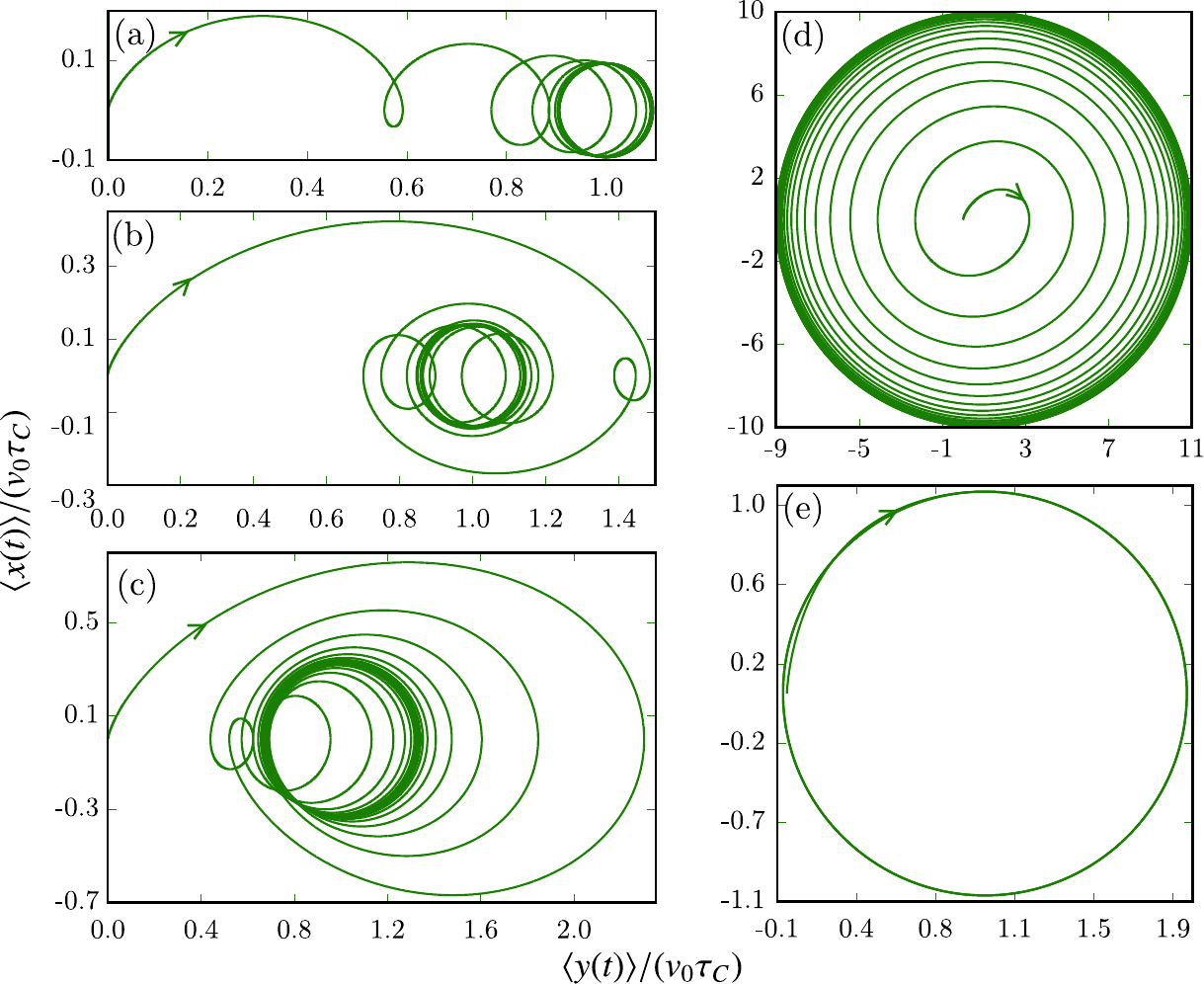}
\caption{Mean displacement (in units of $v_0 \tau_C$) of a jerky active particle with chirality, but in the absence of rotational noise (i.e., $D_r=0$), provided in equation~(\ref{mean_zeroDr}). (a) The timescales are $\tau_J=\tau_C,~\tau_F=2\sqrt{2}\tau_C$ corresponding to $\alpha \approx i0.35/\tau_C$. (b) The timescales are $\tau_J=6 \tau_C,~\tau_F=2\sqrt{2}\tau_C$ corresponding to $\alpha \approx 0.34/\tau_C$. (c) The timescales are $\tau_J=10\tau_C,~\tau_F=2\tau_C$ corresponding to $\alpha \approx 0.5/\tau_C$. (d) The timescales are $\tau_J=10\tau_C,~\tau_F=\tau_C$ corresponding to $\alpha \approx 1/\tau_C$. Note that this is a close to resonance situation where $|\mathrm{Re}[\omega_j]|\approx 1/\tau_C$. (e) The timescales are $\tau_J=0.2\tau_C,~\tau_F=0.2 \tau_C$ corresponding to $\alpha \approx 4.3/\tau_C$.
}
\label{fig: mean_with_chirality_no_noise}
\end{figure}

For vanishing chirality, the mean squared displacement is given as
\begin{eqnarray}
\mathrm{MSD}(t) 
  &=& 2\gamma^2 v_0^2 
        \int_0^t dt'\int_0^{t'} dt''\;e^{-(t'-t'')/\tau_P}\,G(t')\,G(t'') +\;4D\,\gamma^2\int_0^t dt' \; G^2(t').
\end{eqnarray}
We analytically compute the above expression which has the form:
\begin{eqnarray}
\mathrm{MSD}(t) 
  &=& \mathcal{A}_2 t+\mathcal{B}_2+\mathcal{C}_2e^{-t/\tau_P}+\mathcal{D}_2e^{-t/\tau_J}+ e^{-t/(2\tau_J)}\left[\mathcal{E}_2\,\cos(\alpha t)+ \mathcal{F}_2\,\sin(\alpha t)\right]\nonumber\\&&+ e^{-t(1/\tau_P+1/(2\tau_J))}\left[\mathcal{G}_2\,\cos(\alpha t)+ \mathcal{H}_2\,\sin(\alpha t)\right]+e^{-t/\tau_J}\left[\mathcal{I}_2\,\cos(2\alpha t)+ \mathcal{J}_2\,\sin(2\alpha t)\right].\nonumber\\
  \label{msd_zero_omega}
\end{eqnarray}
The prefactors $\{\mathcal{A}_2, ..., \mathcal{J}_2\}$ are complicated functions of the underlying timescales, and their explicit forms are provided in Appendix~\ref{app:app2}. The parameter $\alpha$ is defined in equation~(\ref{alpha}). In the asymptotic limits, we obtain
\begin{eqnarray}
&\mathrm{MSD}(t)& \buildrel t\to0\over=\nonumber \frac{D t^5}{5 \tau_F^4} + \left( \frac{v_0^2}{36 \tau_F^4} - \frac{D}{9 \tau_J \tau_F^4} \right) t^6 + \dots~,\\
&\mathrm{MSD}(t)& \buildrel t\to\infty\over= \left( 4D + {2v_0^2}{\tau_P} \right) t= 4 D_{\mathrm{eff}} t,
\end{eqnarray}
where $D_{\mathrm{eff}}=D+{v_0^2 \tau_P}/{2 }$ is the \emph{effective diffusion constant} for an active particle~\cite{house2007motile,lindner2008diffusion,solon2015active,aragones2018diffusion, jose2022first}. Note that in the long-time limit, the MSD is independent of inertia $m$ and jerk $\lambda$.

Figure~\ref{fig: msd_without_chirality} shows numerical plots of our analytical expression for the mean-squared displacement given in equation~(\ref{msd_zero_omega}). 
The short-time behavior exhibits a super-ballistic regime with $\mathrm{MSD}(t) \sim t^5$, while at long times, the MSD grows linearly in $t$. 
Similar to the mean displacement, the solution displays visible oscillations superimposed on the exponential relaxation for suitable parameter values. These oscillations reflect an underdamped regime, where the system transiently explores rotational modes before relaxing.

\begin{figure}[t!]
\centering
\includegraphics[width=0.9\linewidth]{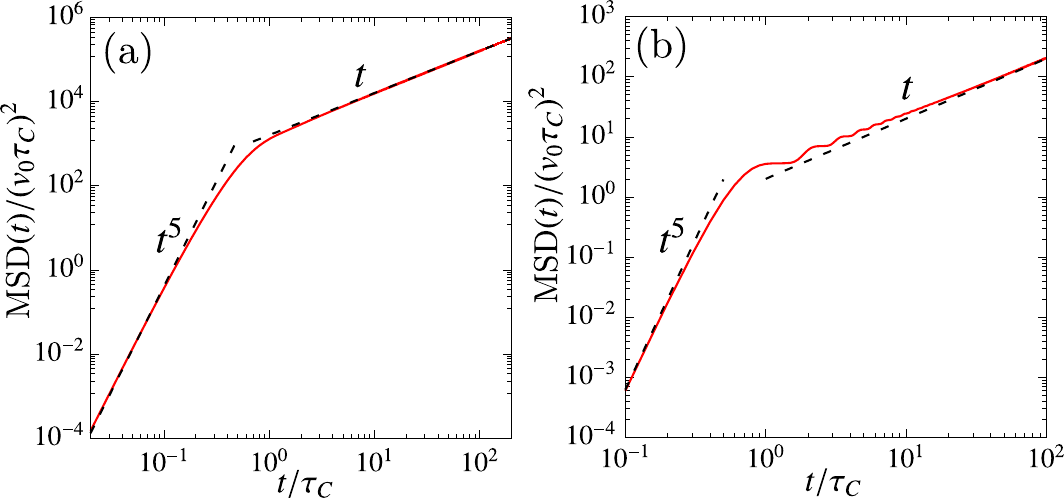}
\caption{Mean squared displacement (in units of ${(v_0 \tau_C)}^2$) as a function of $t/\tau_C$ of a jerky active particle with chirality, but in the absence of rotational noise (i.e., $D_r=0$), provided in equation~(\ref{msd_zero_Dr}). The value $D=0.5$ is kept fixed for both plots. The MSD crosses from $\mathrm{MSD}(t)\propto t^5$ at very short times to $\mathrm{MSD}(t)\propto t$ at large times. (a) The parameter values used are $\tau_J=0.2 \tau_C$, and $\tau_F=0.2 \tau_C$ corresponding to $\alpha \approx 4.3/\tau_C$. (b) The parameter values used are $\tau_J=10 \tau_C$, and $\tau_F=0.4 \tau_C$ corresponding to $\alpha \approx 2.5/\tau_C$. 
}
\label{fig: msd_chirality_zero_noise}
\end{figure}

\section{Effects of finite chirality}
\label{sec:finite_chirality}

\begin{figure}[t]
\centering
\includegraphics[width=1.0\linewidth]{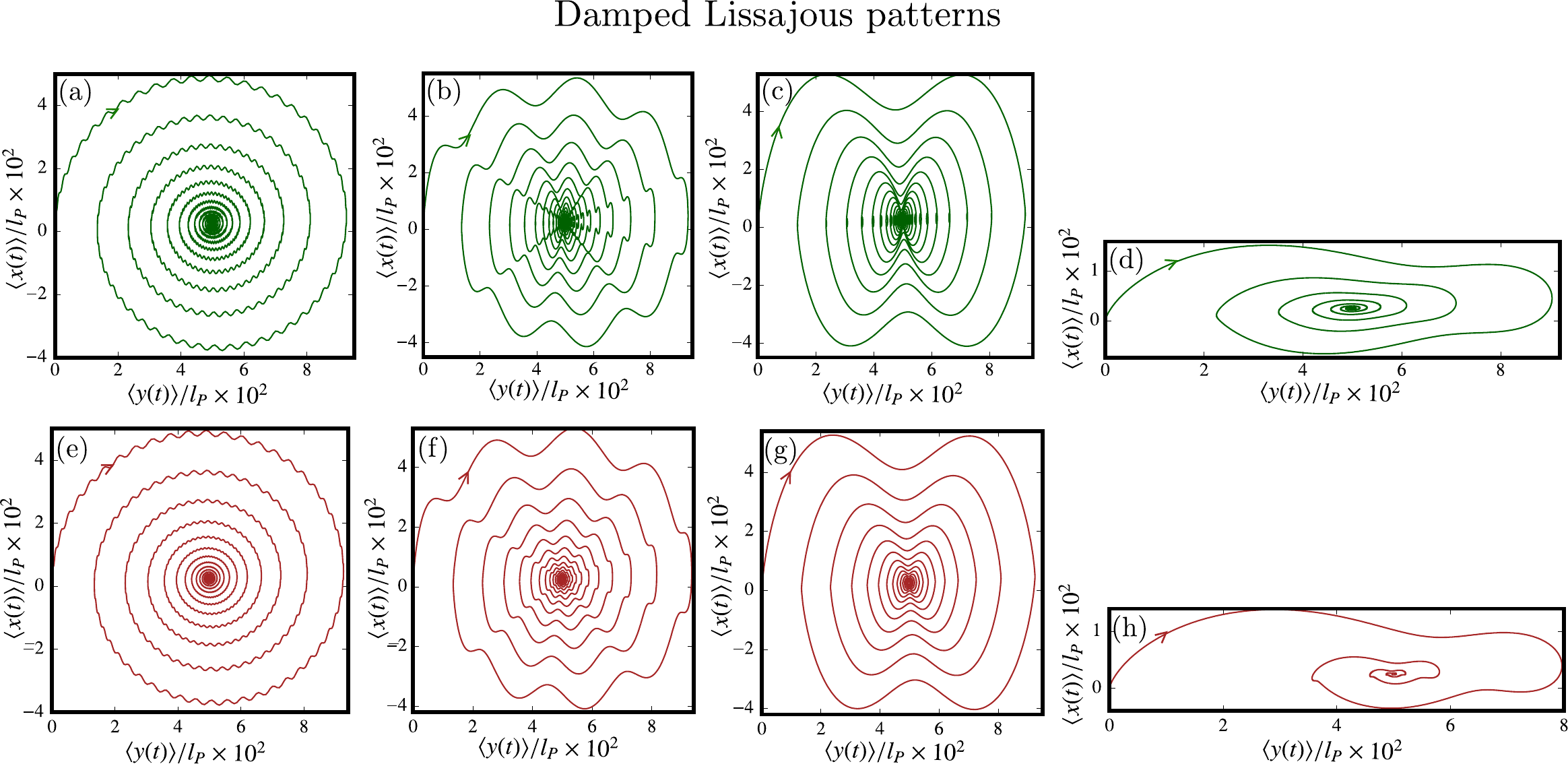}
\caption{\textbf{Mean displacement including initial relaxation when $\tau_P \approx 2 \tau_J$}: Mean displacement in units of $l_P$ as a function of the reduced time $t/\tau_P$ of a jerky active particle with chirality and rotational noise. The parameter values $\tau_P=10,~D=0.5,~v_0=0.05$ and $\tau_C=0.05 \tau_P$ are fixed for all plots. Plots in the same row correspond to same value of $\tau_J$, but $\tau_F$ is varied. (a) $\tau_J=\tau_P, \tau_F=0.001\tau_P$. (b) $\tau_J=\tau_P, \tau_F=0.005\tau_P$. (c) $\tau_J=\tau_P, \tau_F=0.01\tau_P$. (d)~$\tau_J=\tau_P, \tau_F=0.2\tau_P$. (e) $\tau_J=0.5\tau_P, \tau_F=0.001\tau_P$. (f) $\tau_J=0.5\tau_P, \tau_F=0.005\tau_P$. (g) $\tau_J=0.5\tau_P, \tau_F=0.01\tau_P$. (h) $\tau_J=0.5\tau_P, \tau_F=0.2\tau_P$. We observe disortions and beats on the spirals.
}
\label{fig: mean_with_chirality_and_noise_tauf}
\end{figure}

We next analyze the case where the particle has a systematic rotational drift (chirality). Let us first consider the noiseless case where $D_r=0$. 
\subsection{Case 1:  Zero rotational noise}

The mean displacement is given as
\begin{eqnarray}
    \langle \vec{r}(t) \rangle = \gamma v_0 \int_0^t dt' G(t - t')
\left(\begin{array}{c}
\cos(t'/\tau_C) \\
\sin(t'/\tau_C)
\end{array}\right).
\end{eqnarray}
We compute the analytical expressions for $x(t)$ and $y(t)$ by substituting the expression for the Green's function provided in equation~(\ref{eqn:greens_fn}). This yields
\begin{eqnarray}
\langle x(t) \rangle &=& \nonumber
\frac{v_0}{\tau_F^2 \prod_{j=1}^2 (\omega_j^2 - 1/\tau_C^2)}  \Big[
\frac{i  e^{-i t \sum_{j=1}^2 \omega_j}}{\sum_{j=1}^2 (-1)^j \omega_j}  \sum_{j=1}^2 (-1)^j e^{i t \omega_j} (\omega_j^2 - 1/\tau_C^2)
\\&&- \nonumber i \sum_{j=1}^2 \omega_j \cos(t/\tau_C)
- \tau_C{(\prod_{j=1}^2 \omega_j + 1/\tau_C^2)} \sin(t/\tau_C)
\Big],\\
\langle y(t) \rangle &=&\nonumber \frac{v_0}{\tau_F^2 \prod_{j=1}^2 (  \omega_j^2-1/\tau_C^2)} \Big[
\nonumber \frac{e^{-i t \sum_{j=1}^2 \omega_j}}{\tau_C\sum_{j=1}^2(-1)^{j}\omega_j\prod_{j=1}^2\omega_j}
\sum_{j=1}^2 (-1)^{j} 
{e^{i t\omega_j }\omega_j(\omega_j^2 -  1/\tau_C^2)} \\
&& 
- \tau_C{(\prod_{j=1}^2 \omega_j + 1/\tau_C^2)} \cos( t/\tau_C) 
+ i \sum_{j=1}^2 \omega_j  \sin( t/\tau_C)+\frac{\tau_C\prod_{j=1}^2 ( \omega_j^2-1/\tau_C^2 )}{\prod_{j=1}^2 \omega_j} 
\Big],
\end{eqnarray}
where $\sum_{j=1}^2 \omega_j={-i}/{\tau_J}$,~$\sum_{j=1}^2(-1)^{j}\omega_j=-2\sqrt{{ {1}/{ \tau_F^2}} - { {1}/{(4\tau_J^2)}}}$~and~$\prod_{j=1}^2 \omega_j=-{1}/{\tau_F^2}$.
After substituting the expressions for the eigen-frequencies provided in equation~(\ref{omega_j}), 
we obtain the compact forms:
\begin{eqnarray}
\langle x(t) \rangle &=& \mathcal{A}_3\,\cos( t/\tau_C)+ \mathcal{B}_3\,\sin( t/\tau_C)+e^{-t/(2\tau_J)}\left[\mathcal{C}_3\,\cos(\alpha t)+ \mathcal{D}_3\,\sin(\alpha t)\right]
,\nonumber\\
\langle y(t) \rangle &=& v_0\tau_C-\mathcal{B}_3\,\cos( t/\tau_C)+ \mathcal{A}_3\,\sin( t/\tau_C)+e^{-t/(2\tau_J)}\left[\mathcal{E}_3\,\cos(\alpha t)+ \mathcal{F}_3\,\sin(\alpha t)\right].
\label{mean_zeroDr}
\end{eqnarray}
The prefactors $\{\mathcal{A}_3, ..., \mathcal{F}_3\}$ are complicated functions of the underlying timescales, and their explicit forms are provided in Appendix~\ref{app:app2}.
The above expressions have the limiting behaviors
\begin{eqnarray}
&\langle x(t) \rangle& 
\buildrel t\to0\over= \nonumber
 \frac{v_0 t^3}{6 \tau_F^2} - \frac{v_0  t^4}{24 \tau_J \tau_F^2}+ \dots~,\\
&\langle y(t) \rangle& 
\buildrel t\to0\over= 
 \frac{v_0  t^4}{24 \tau_F^2 \tau_C} -\frac{{v_0}  t^5}{120 \tau_J  \tau_F^2 \tau_C }+ \dots~.
\end{eqnarray}
After the initial transient regime driven by jerk, inertia and damping, the trajectory converges to a steady circular orbit at large times:
\begin{eqnarray}
&\langle x(t) \rangle& \buildrel t\to \infty\over= \nonumber x_c + r \cos\left(t/\tau_C + \phi\right),\\ 
&\langle y(t) \rangle& \buildrel t\to \infty\over= y_c + r \sin\left(t/\tau_C + \phi\right),
\end{eqnarray}
with center 
\begin{equation}
(x_c,~y_c)=\left(0,{v_0 \tau_C}\right),
\end{equation}
radius 
\begin{equation}
   r=\sqrt{\mathcal{A}_3^2+\mathcal{B}_3^2}= \frac{v_0 \tau_C}{  \sqrt{{{\tau_F}{}^4 }/{({\tau_J}{}^2\tau_C
   ^2)}+\left({{\tau_F}{}^2} /{\tau_C
   ^2}-1\right)^2}},
   \label{radius}
\end{equation}
phase shift
\begin{equation}
\phi=\tan^{-1}(-\mathcal{B}_3/\mathcal{A}_3)=\tan^{-1}\left[\tau _J \left({\tau _C}/{\tau_F^2}-{1}/{\tau _C}\right)\right],  
\end{equation}
and frequency $\omega_0=1/\tau_C$. We note that the above expressions are non-trivial functions of jerk, friction and chirality. It is well-known that chiral active particles in the absence of rotational noise exhibit circular swimming behavior at large times~\cite{teeffelen2008dynamics,sprenger2021rocket}. The radius of the orbit is determined by the underlying parameters. Equation~(\ref{radius}) in the $\lambda \rightarrow 0$ limit yields the known expression $r=v_0 \gamma/(\omega_0\sqrt{m^2 \omega_0^2+\gamma^2})$ for an underdamped ABP with inertia~\cite{sprenger2021rocket} and further in the overdamped limit $m \rightarrow 0$, one obtains $r=v_0/ \omega_0$. Figure~\ref{fig: mean_with_chirality_no_noise} shows the numerical plots of the noise‐free analytic expressions for the mean displacements $\langle x(t)\rangle$ and $\langle y(t)\rangle$ in the presence of chirality, given in~equation~(\ref{mean_zeroDr}). For different parameter choices, all trajectories converge to a circle at long times. However, the initial relaxation is complex and strongly dependent on the parameters. 
\begin{figure}[t!]
\centering
\includegraphics[width=1.0\linewidth]{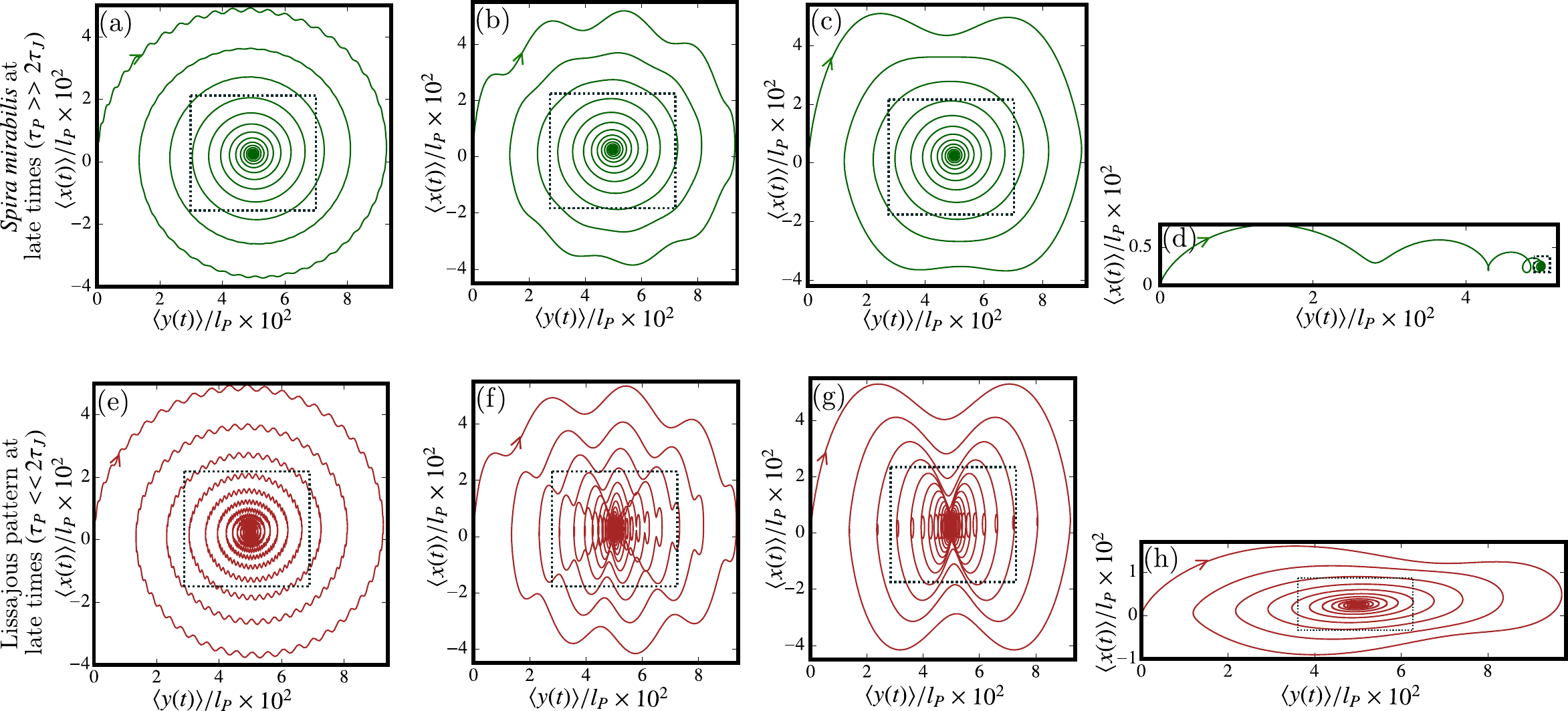}
\caption{\textbf{\emph{Spira mirabilis} vs. damped Lissajous patterns at large times}:~Mean displacement in units of $l_P$ as a function of the reduced time $t/\tau_P$ of a jerky active particle with chirality and rotational noise. The parameter values $\tau_P=10,~D=0.5,~v_0=0.05$ and $\tau_C=0.05 \tau_P$ are fixed for all plots. Plots in the same row correspond to same value of $\tau_J$, but $\tau_F$ is varied. \textbf{Mean displacement including initial relaxation when $\tau_P >> 2 \tau_J$}: (a) $\tau_J=0.1\tau_P, \tau_F=0.001\tau_P$. (b) $\tau_J=0.1\tau_P, \tau_F=0.005\tau_P$. (c) $\tau_J=0.1\tau_P, \tau_F=0.01\tau_P$. (d)~$\tau_J=0.1\tau_P, \tau_F=0.2\tau_P$. The trajectories converge to a \emph{spira mirabilis} at large times. \textbf{Mean displacement including initial relaxation when $\tau_P << 2 \tau_J$}: (e) $\tau_J=2\tau_P, \tau_F=0.001\tau_P$. (f) $\tau_J=2\tau_P, \tau_F=0.005\tau_P$. (g) $\tau_J=2\tau_P, \tau_F=0.01\tau_P$. (h) $\tau_J=2\tau_P, \tau_F=0.2\tau_P$. The trajectories converge to a Lissajous pattern at large times. 
}
\label{fig: mean_with_chirality_and_noise_convergence}
\end{figure}

The mean squared displacement is given as
\begin{eqnarray}
\mathrm{MSD}(t) 
  &=& 2\gamma^2 v_0^2 
        \int_0^t dt'\int_0^{t'} dt''\;\cos\!\bigl[(t'-t'')/ \tau_C\bigr]\,G(t')\,G(t'')+\;4D\,\gamma^2\int_0^t dt' \; G^2(t').
\label{msd_zero_Dr}
\end{eqnarray}
Since the explicit expression for the MSD is quite large, we list the limiting behaviors
\begin{eqnarray}
&\mathrm{MSD}(t)& \buildrel t\to0\over= \nonumber \frac{D t^5}{5 \tau_F^4} +  \frac{v_0^2  t^6}{36 \tau_F^4}  + \dots~,\\
&\mathrm{MSD}(t)& \buildrel t\to\infty\over=  4 {D} t.
\end{eqnarray}
The full dynamical behavior of MSD for two different choices of parameter values is shown in figure~\ref{fig: msd_chirality_zero_noise}. Similar to the case of vanishing chirality discussed before, we observe smooth relaxation or oscillatory behavior of MSD depending on the relative magnitude of the timescales.


\subsection{Case 1:  Non-zero rotational noise}
\label{Dr_w}
In this section, we consider the effects of finite rotational noise on the dynamics of a jerky chiral active particle. The mean displacement is given by equation~(\ref{eq:MD}). After computing the integral, we obtain
\begin{eqnarray}
\langle x(t) \rangle &=& \frac{v_0 }{\tau_P(\frac{1}{\tau_P^2}+\frac{1}{\tau_C^2})}+e^{-t/\tau_P}\left[\mathcal{A}_4\,\cos( t/\tau_C)+ \mathcal{B}_4\,\sin( t/\tau_C)\right]\nonumber\\&&+e^{-t/(2\tau_J)}\left[\mathcal{C}_4\,\cos(\alpha t)+ \mathcal{D}_4\,\sin(\alpha t)\right]
,\nonumber\\
\langle y(t) \rangle &=& \frac{v_0 }{\tau_P(\frac{1}{\tau_P^2}+\frac{1}{\tau_C^2})}+e^{-t/\tau_P}\left[-\mathcal{B}_4\,\cos( t/\tau_C)+ \mathcal{A}_4\,\sin( t/\tau_C)\right]\nonumber\\&&+e^{-t/(2\tau_J)}\left[\mathcal{E}_4\,\cos(\alpha t)+ \mathcal{F}_4\,\sin(\alpha t)\right].
\label{mean_Dr_chirality}
\end{eqnarray}
The prefactors $\{\mathcal{A}_4, ..., \mathcal{F}_4\}$ are complicated functions of the underlying timescales, and their explicit forms are provided in Appendix~\ref{app:app2}.
The above expressions have the limiting behaviors
\begin{eqnarray}
\langle x(t) \rangle 
&\buildrel t\to0\over=& 
 \frac{v_0 t^3}{6 \tau_F^2} - \frac{v_0  }{24  \tau_F^2}\left(\frac{1}{\tau_J}+\frac{1}{\tau_P}\right)t^4+ \dots~,\nonumber\\
\langle y(t) \rangle 
&\buildrel t\to0\over=& 
 \frac{v_0  t^4}{24 \tau_F^2 \tau_C} -\frac{{v_0} }{60  \tau_F^2 \tau_C}\left(\frac{1}{2\tau_J }+\frac{1}{\tau_P}\right)t^5+ \dots~.
\end{eqnarray}

\begin{figure}[t!]
\centering
\includegraphics[width=1.0\linewidth]{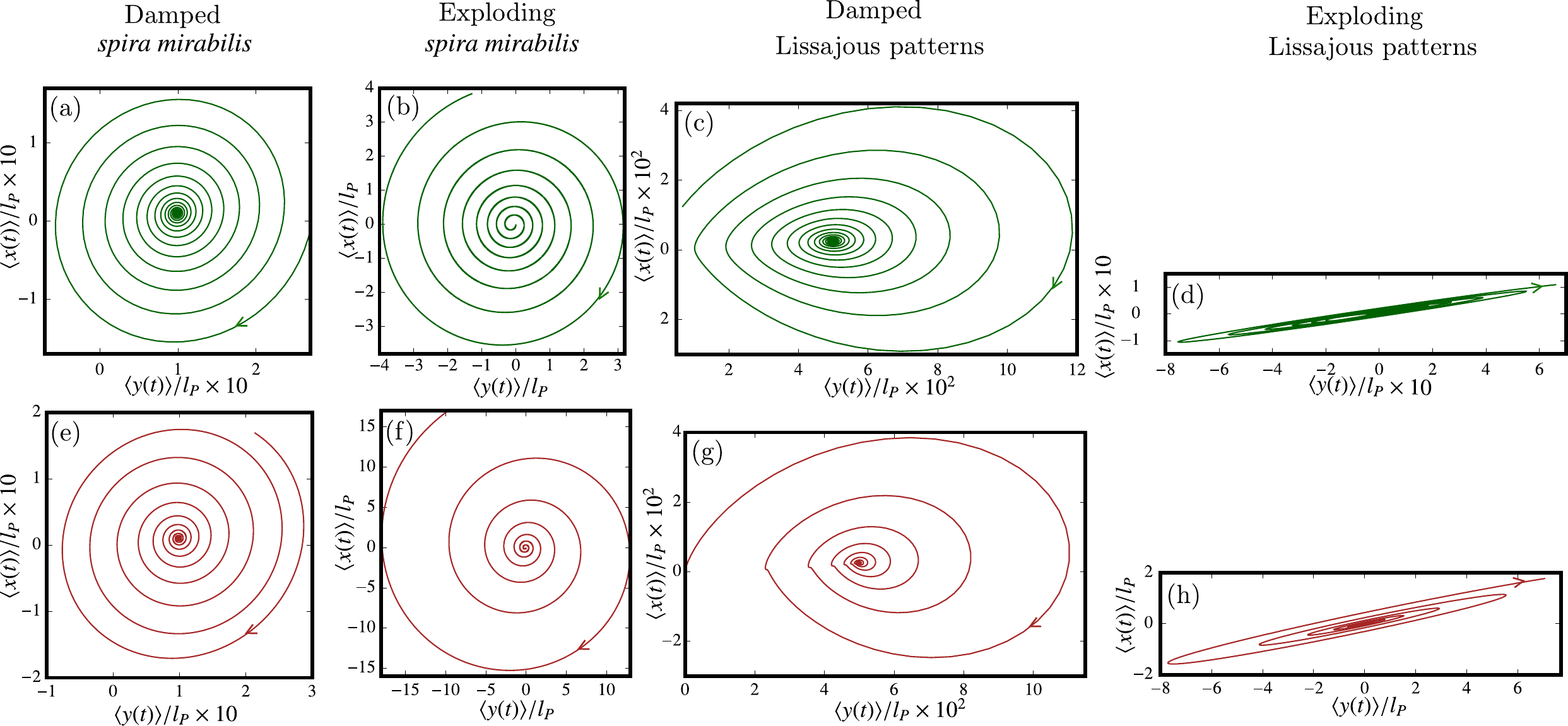}
\caption{\textbf{Mean displacement at late times}:~Mean displacement in units of $l_P$ as a function of the reduced time $t/\tau_P$ of a jerky active particle with chirality and rotational noise. The parameter values $\tau_P=10,~D=0.5,~v_0=0.05$ and $\tau_F=0.1 \tau_P$ are fixed for all plots. Plots in the same row correspond to same value of $|\tau_J|$, but $\tau_C$ is varied. (a) $\tau_J=\tau_P,~\tau_C=0.1\tau_P$. (b)~$\tau_J=-\tau_P,~\tau_C=0.1\tau_P$. (c)~$\tau_J=\tau_P,~\tau_C=0.05\tau_P$. (d)~$\tau_J=-\tau_P,~\tau_C=0.05\tau_P$. (e) $\tau_J=0.5\tau_P,~\tau_C=0.1\tau_P$. (f)~$\tau_J=-0.5\tau_P,~\tau_C=0.1\tau_P$. (g)~$\tau_J=0.5\tau_P,~\tau_C=0.05\tau_P$. (h)~$\tau_J=-0.5\tau_P,~\tau_C=0.05\tau_P$. We observe damped or exploding spirals depending on the sign of $\tau_J$.
}
\label{fig: mean_with_chirality_and_noise}
\end{figure}

At long times, one observes complex trajectory patterns with distortions and oscillations depending on the choice of parameters (see figures~\ref{fig: mean_with_chirality_and_noise} and~\ref{fig: mean_with_chirality_and_noise_tauf}). The centre of the spiral is given as
\begin{equation}
(x_c^s,~y_c^s)=\left(\frac{v_0 }{\tau_P(\frac{1}{\tau_P^2}+\frac{1}{\tau_C^2})},\frac{v_0 }{\tau_C(\frac{1}{\tau_P^2}+\frac{1}{\tau_C^2})}\right),
\end{equation}
and is independent of the values of $\tau_J$ and $\tau_F$. Equation~(\ref{mean_Dr_chirality}) can also be rewritten as
\begin{eqnarray}
\langle x(t) \rangle  &=& \nonumber x_c^s+r_1 e^{-t/\tau_P} \cos( t/\tau_C + \phi_1) + r_2 e^{- t/(2\tau_J)} \cos(\alpha t + \phi_2), \\
\langle y(t) \rangle &=&   y_c^s+r_1 e^{-t/\tau_P} \sin(t/\tau_C + \phi_1) + r_3 e^{- t/(2 \tau_J)} \cos(\alpha t + \phi_3).
\label{mean_lissajous}
\end{eqnarray}
Equation~(\ref{mean_lissajous}) is a central result of this paper. The constants $r_1=\sqrt{\mathcal{A}_4^2+\mathcal{B}_4^2}$, $r_2=\sqrt{\mathcal{C}_4^2+\mathcal{D}_4^2}$, $r_3=\sqrt{\mathcal{E}_4^2+\mathcal{F}_4^2}$ and the angles $\phi_1=\tan^{-1}(-\mathcal{B}_4/\mathcal{A}_4)$, $\phi_2=\tan^{-1}(-\mathcal{D}_4/\mathcal{C}_4)$, $\phi_3=\tan^{-1}(\mathcal{E}_4/\mathcal{F}_4)$ depend on the values of the underlying parameters. The position vector $\vec{r}(t)$ is given by the superposition of two exponentially decaying oscillatory modes:
\begin{equation}
\langle \vec{r}(t) \rangle = \vec{A}(t) + \vec{B}(t),
\end{equation}
where $\vec{A}(t)$ traces a circular (logarithmic) spiral, and $\vec{B}(t)$ forms a distorted elliptical spiral (a Lissajous type curve), each characterized by distinct frequencies $1/\tau_C=\omega_0,~\alpha$ and decay rates $\tau_P,~2\tau_J$ respectively. Note that $\vec{B}(t)$ is not a circular spiral as $r_2 \neq r_3$ and $\phi_2 \neq \phi_3$. The  interference of $\vec{A}(t)$ and $\vec{B}(t)$ generates a complex spiral trajectory.  The trajectory becomes distorted and exhibit lobed or beat-like patterns, depending on the interference between the two components (see figure~\ref{fig: mean_with_chirality_and_noise_tauf}). As time progresses, the more slowly decaying mode dominates, and the trajectory asymptotically approaches the corresponding spiral. i.e., if $\tau_P >> 2\tau_J$, $\vec{r}(t) \rightarrow \vec{A}(t)$ and if $\tau_P << 2\tau_J$, $\vec{r}(t) \rightarrow \vec{B}(t)$ (see figure~\ref{fig: mean_with_chirality_and_noise_convergence}). Note that although logarithmic spirals are self-similar~\cite{babel2014swimming}, Lissajous curves are not in general self similar. The value of $\tau_J$ can be positive or negative depending on the sign of $\lambda$ and $m$. For $\tau_J>0$ (stable case) the trajectory spirals inward (damped spirals) while for $\tau_J<0$ (unstable case), the trajectories spiral outwards (exploding spirals). The direction of the spirals is set solely by the sign of $\tau_J$~(see figure~\ref{fig: mean_with_chirality_and_noise}).

\begin{figure}[t!]
\centering
\includegraphics[width=0.9\linewidth]{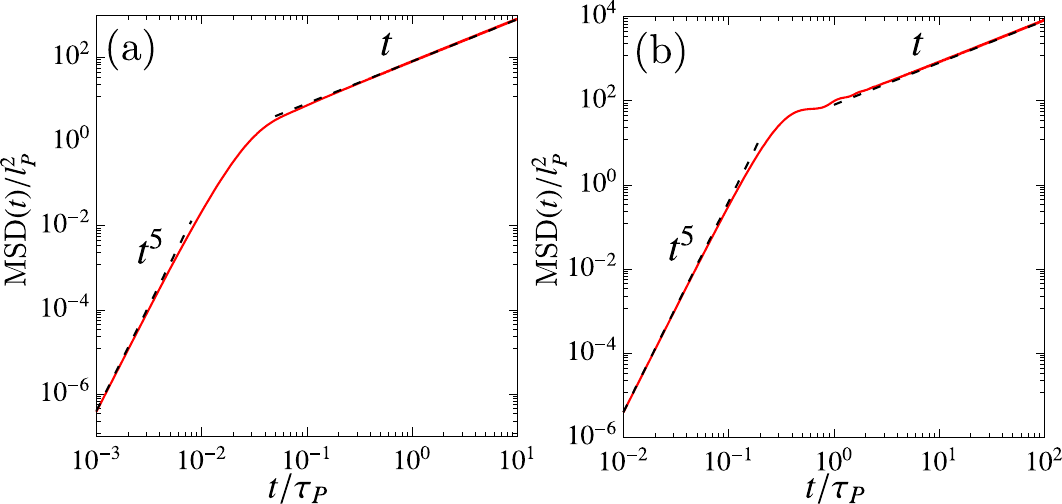}
\caption{Mean squared displacement of a jerky active particle with both chirality and rotational diffusion given in equation~(\ref{eq:msd}). We have set $\tau_P=10,~v_0=0.05$,~and~$\tau_C=0.5$. The MSD crosses from $\mathrm{MSD}(t)\propto t^5$ at very short times to $\mathrm{MSD}(t)\simeq4 D_c t$ at large times.~(a)~The parameter values used are $\tau_J=0.01 \tau_P$,~and~$\tau_F=0.01 \tau_P$.~(b)~The parameter values used are $\tau_J=0.5 \tau_P$,~and~$\tau_F=0.1\tau_P$. 
}
\label{fig: msd_chirality}
\end{figure}

The mean squared displacement is given by equation~(\ref{eq:msd}). Since this expression is quite large, we list the limiting behaviors
\begin{equation}
\mathrm{MSD}(t) \buildrel t\to0\over= \frac{D t^5}{5 \tau_F^4} + \left( \frac{v_0^2}{36 \tau_F^4} - \frac{D}{9 \tau_J \tau_F^4} \right) t^6 + \dots~,
\end{equation}
and
\begin{equation}
\mathrm{MSD}(t) \buildrel t\to\infty\over= \left( 4D + \frac{2v_0^2}{\left(\frac{1}{\tau_P}+\frac{\tau_P}{\tau_C^2} \right)} \right) t= 4 D_{\mathrm{c}} t,
\end{equation}
where $D_{\mathrm{c}}=D+\frac{v_0^2}{2 \left({1}/{\tau_P}+{\tau_P}/{\tau_C^2} \right)}$ is the \emph{effective diffusion constant} for an active particle with chirality~\cite{teeffelen2008dynamics}. The full dynamical behavior of MSD for two different choices of parameter values is shown in figure~\ref{fig: msd_chirality}. Similar to the cases discussed before, we observe smooth relaxation or oscillatory behavior of MSD depending on the relative magnitude of the timescales.

\section{Conclusions and discussion}
\label{sec:conclusion}
In this work, we introduced a new class of active particles that incorporate both jerk and chirality, namely a jerky chiral active Brownian particle (jcABP) model. Understanding the behavior of a single active particle with jerk provides valuable insight into the roles of memory and delayed response in active systems.  Interestingly, we found that jerk not only modifies the standard circular swimming behavior of chiral active particles but also gives rise to complex spiral trajectories and oscillatory patterns, revealing new dynamical features. This model can be naturally generalized to higher spatial dimensions and to many-particle systems. For example, models of interacting active particles display rich collective phenomena, including motility-induced phase separation (MIPS), clustering, and flocking. At the continuum level, hydrodynamic theories have been developed to describe polar and nematic order in dense active fluids. More recent studies have also investigated activity-induced pattern formation in confined geometries, responses to external fields, and collective dynamics in chiral systems. Our work lays the foundation for incorporating memory effects into these broader theoretical frameworks.

\ack{S.J. acknowledges support from the Alexander von Humboldt Foundation through a Humboldt Research Fellowship. H.L. acknowledges funding of the German Research Foundation (DFG) project LO 418/29-1.}


\appendix               
\setcounter{section}{0}

\section{Orientation correlations}\label{app:app1}
 Integrating equation~(\ref{eq:theta_dyn}) gives the angular displacement:
\begin{equation}
\theta(t) = \omega_0 t + \sqrt{2D_r} \int_0^t dt'\;\eta_r(t') .
\end{equation}
Since \( \theta(t) \) is a Gaussian process with mean \( \langle \theta(t) \rangle = \omega_0 t \) and variance
\begin{equation}
\langle [\theta(t) - \langle \theta(t) \rangle]^2 \rangle_c = 2D_r t,
\end{equation}
we can use the identity for Gaussian averages:
\begin{equation}
\langle e^{i \theta(t)} \rangle = e^{i \langle \theta(t) \rangle} e^{ -\frac{1}{2} \langle [\theta(t) - \langle \theta(t) \rangle]^2 \rangle_c } = e^{i \omega_0 t} e^{-D_r t}.
\end{equation}
Separating real and imaginary parts gives
\begin{eqnarray}
\langle \cos \theta(t) \rangle &= e^{-D_r t} \cos(\omega_0 t), \\
\langle \sin \theta(t) \rangle &= e^{-D_r t} \sin(\omega_0 t).
\end{eqnarray}
Thus, the average orientation vector is
\begin{equation}
\langle \hat{{n}}(t) \rangle = e^{-D_r t}
\left(\begin{array}{c}
\cos(\omega_0 t) \\
\sin(\omega_0 t)
\end{array}\right).
\end{equation}
To compute the orientation autocorrelation function, we consider
\begin{eqnarray}
\langle \hat{n}(t_1) \cdot \hat{n}(t_2) \rangle &= \langle \cos[\theta(t_1) - \theta(t_2)] \rangle.
\end{eqnarray}
Let \( \Delta t = |t_1 - t_2| \). Then,
\begin{eqnarray}
\theta(t_1) - \theta(t_2) &= \omega_0 \Delta t + \sqrt{2D_r} \int_{t_2}^{t_1} dt' \; \eta_r(t') ,
\end{eqnarray}
and therefore
\begin{eqnarray}
\langle \cos[\theta(t_1) - \theta(t_2)] \rangle 
&=&\nonumber \langle \cos\left( \omega_0 \Delta t + \sqrt{2D_r} \int_{t_2}^{t_1} dt' \; \eta_r(t')  \right) \rangle \\
&=& \cos(\omega_0 \Delta t)~e^{-D_r \Delta t}.
\end{eqnarray}
This gives the orientation correlation function given in equation~(\ref{orientation_correlation}).
\section{Expressions for prefactors}\label{app:app2}

\begin{eqnarray}
\mathcal{A}_1&=&\frac{\tau _J \tau _P^2}{\tau _J \tau _P^2+\tau _F^2 \left(\tau _J-\tau _P\right)},\nonumber\\ \mathcal{B}_1&=&-\frac{2 \tau _J^2 \tau _P+\tau _F^2 \left(\tau _J-\tau _P\right)}{2 \alpha  \tau _J
   \left(\tau _F^2 \left(\tau _J-\tau _P\right)+\tau _J \tau _P^2\right)}, \nonumber\\ \mathcal{C}_1&=&-\frac{\tau _F^2 \left(\tau _J-\tau _P\right)}{\tau _F^2 \left(\tau _J-\tau
   _P\right)+\tau _J \tau _P^2}.
   \end{eqnarray}

\begin{eqnarray}
\mathcal{A}_2&=&4 D+2 v_0^2 \tau _P,\nonumber\\ \mathcal{B}_2&=&\frac{2 D \left(-3 \tau _F^2+\tau _J^2\right)}{\tau _J}-\frac{v_0^2 \tau _P \left(2 \tau _J^2 \tau _P^3+3
   \tau _F^4 \left(\tau _J+\tau _P\right)-\tau _F^2 \left(\tau _J^3-4 \tau _J \tau
   _P^2\right)\right)}{\tau _J^2 \tau _P^2+\tau _F^2 \tau _J \left(\tau _J+\tau _P\right)}, \nonumber\\ \mathcal{C}_2&=&\frac{2 v_0^2 \tau _J \tau _P^4}{\tau _F^2 \left(\tau _J-\tau _P\right)+\tau _J \tau _P^2},\nonumber\\ \mathcal{D}_2&=& \frac{2 \tau _J^3 \left(4 D \tau _J \tau _P^2+\tau _F^2 \left(2 \tau _J \left(2 D+v_0^2 \tau
   _P\right)-\tau _P \left(4 D+v_0^2 \tau _P\right)\right)\right)}{\left(\tau _F^2-4 \tau _J^2\right)
   \left(\tau _F^2 \left(\tau _J-\tau _P\right)+\tau _J \tau _P^2\right)},\nonumber\\ \mathcal{E}_2&=&\frac{4 \tau _F^4 \left(\tau _J-\tau _P\right) \left(2 D+v_0^2 \tau _P\right)+2 \tau _F^2 \tau _J \tau
   _P^2 \left(4 D+v_0^2 \left(\tau _J+\tau _P\right)\right)}{\tau _J \left(\tau _F^2 \left(\tau _J-\tau
   _P\right)+\tau _J \tau _P^2\right)},
    \nonumber\\ \mathcal{F}_2&=&\frac{4 \tau _F^5 \left(\tau _J-\tau _P\right) \left(2 D+v_0^2 \tau _P\right)-4 \tau _F \tau _J^3 \tau
   _P^2 \left(4 D+v_0^2 \tau _P\right)+2 \tau _F^3 \tau _J \left(-4 \tau _J^2 \left(2 D+v_0^2 \tau
   _P\right)+\tau _P^2 \left(4 D+v_0^2 \tau _P\right)+\tau _J \tau _P \left(8 D+5 v_0^2 \tau
   _P\right)\right)}{\tau _J \sqrt{-\tau _F^2+4 \tau _J^2} \left(\tau _F^2 \left(\tau _J-\tau
   _P\right)+\tau _J \tau _P^2\right)},
     \nonumber\\ \mathcal{G}_2&=&-\frac{2 v_0^2 \tau _F^2 \tau _J \tau _P^4 \left(\tau _J+\tau _P\right)}{2 \tau _F^2 \tau _J^2 \tau
   _P^2+\tau _J^2 \tau _P^4+\tau _F^4 \left(\tau _J^2-\tau _P^2\right)},
      \nonumber\\ \mathcal{H}_2&=&-\frac{2 v_0^2 \tau _F \tau _J \tau _P^4 \left(-2 \tau _J^2 \tau _P+\tau _F^2 \left(\tau _J+\tau
   _P\right)\right)}{\sqrt{-\tau _F^2+4 \tau _J^2} \left(2 \tau _F^2 \tau _J^2 \tau _P^2+\tau _J^2 \tau
   _P^4+\tau _F^4 \left(\tau _J^2-\tau _P^2\right)\right)},
       \nonumber\\ \mathcal{I}_2&=&
       -\frac{\tau _F^2 \left(v_0^2 \left(2 \tau _J^4+\tau _F^4 \left(1-\frac{\tau _J}{\tau _P}\right)+\tau _F^2
   \tau _J^2 \left(-4+\frac{3 \tau _J}{\tau _P}\right)\right)+2 D \left(-\tau _F^2+3 \tau _J^2\right)
   \left(\tau _J+\frac{\tau _F^2 \left(\tau _J-\tau _P\right)}{\tau _P^2}\right)\right)}{\tau _J
   \left(-\tau _F^2+4 \tau _J^2\right) \left(\tau _J+\frac{\tau _F^2 \left(\tau _J-\tau _P\right)}{\tau
   _P^2}\right)}, \nonumber\\ \mathcal{J}_2&=&\frac{2 D \tau _F \tau _J^3 \tau _P^2-\tau _F^5 \left(\tau _J-\tau _P\right) \left(2 D+v_0^2 \tau
   _P\right)+\tau _F^3 \tau _J \left(-2 D \tau _P^2-2 \tau _J \tau _P \left(D+v_0^2 \tau _P\right)+\tau
   _J^2 \left(2 D+v_0^2 \tau _P\right)\right)}{\tau _J \sqrt{-\tau _F^2+4 \tau _J^2} \left(\tau _F^2
   \left(\tau _J-\tau _P\right)+\tau _J \tau _P^2\right)}.\nonumber\\
   \end{eqnarray}

  \begin{eqnarray}
\mathcal{A}_3&=&-\frac{v_0 \tau _F^2 \tau _J}{\frac{\tau _F^4}{\tau _C^2}+\left(-1+\frac{\tau _F^2}{\tau _C^2}\right)^2
   \tau _J^2},\nonumber\\ \mathcal{B}_3&=&\frac{v_0 \tau _C^3 \left(\tau _C^2-\tau _F^2\right) \tau _J^2}{\tau _C^2 \tau _F^4+\left(\tau _C^2-\tau
   _F^2\right)^2 \tau _J^2}, \nonumber\\ \mathcal{C}_3&=&\frac{v_0 \tau _F^2 \tau _J}{\frac{\tau _F^4}{\tau _C^2}+\left(-1+\frac{\tau _F^2}{\tau _C^2}\right)^2
   \tau _J^2}, \nonumber\\ \mathcal{D}_3&=&\frac{v_0 \tau _F \tau _J \left(\tau _F^2+2 \left(-1+\frac{\tau _F^2}{\tau _C^2}\right) \tau
   _J^2\right)}{\sqrt{-\tau _F^2+4 \tau _J^2} \left(\frac{\tau _F^4}{\tau _C^2}+{\left(-1+\frac{\tau
   _F^2}{\tau _C^2}\right)}^2 \tau _J^2\right)},\nonumber\\ \mathcal{E}_3&=&
\frac{v_0 \tau _F^2 \left(\tau
   _F^2+\left(-1+\frac{\tau _F^2}{\tau
   _C^2}\right) \tau _j^2\right)}{\tau _C
   \left(\frac{\tau _F^4}{\tau
   _C^2}+\left(-1+\frac{\tau _F^2}{\tau
   _C^2}\right)^2 \tau _j^2\right)},\nonumber\\ \mathcal{F}_3&=&\frac{v_0 \tau _F^3 \left(\tau _F^2+\left(-3+\frac{\tau _F^2}{\tau _C^2}\right) \tau _j^2\right)}{\tau _C
   \sqrt{-\tau _F^2+4 \tau _j^2} \left(\frac{\tau _F^4}{\tau _C^2}+\left(-1+\frac{\tau _F^2}{\tau
   _C^2}\right)^2 \tau _j^2\right)}.
   \end{eqnarray}

   \begin{eqnarray}
\mathcal{A}_4&=&-\frac{v_0 \tau _C^4 \tau _F^2 \tau _P^3 \left(\tau _P^2 \left(-3+\tau _J \tau _P\right)+\tau _C^2
   \left(1-\tau _J \tau _P+\tau _F^2 \tau _P^2\right)\right)}{\left(\tau _C^2+\tau _P^2\right) \left(\tau
   _P^4+\tau _C^4 \left(1-\tau _J \tau _P+\tau _F^2 \tau _P^2\right){}^2+\tau _C^2 \tau _P^2 \left(2-2
   \tau _J \tau _P+\left(-2 \tau _F^2+\tau _J^2\right) \tau _P^2\right)\right)},\nonumber\\ \mathcal{B}_4&=&\frac{v_0 \tau _C^3 \tau _F^2 \tau _P^4 \left(-\tau _P^2+\tau _C^2 \left(3-2 \tau _J \tau _P+\tau _F^2
   \tau _P^2\right)\right)}{\left(\tau _C^2+\tau _P^2\right) \left(\tau _P^4+\tau _C^4 \left(1-\tau _J
   \tau _P+\tau _F^2 \tau _P^2\right){}^2+\tau _C^2 \tau _P^2 \left(2-2 \tau _J \tau _P+\left(-2 \tau
   _F^2+\tau _J^2\right) \tau _P^2\right)\right)}, \nonumber\\ \mathcal{C}_4&=&\frac{v_0 \tau _C^2 \tau _P \left(-\tau _P^2-\tau _C^2 \left(-1+\tau _J \tau _P\right) \left(-1+\tau _P
   \left(\tau _J-\tau _F^2 \tau _P\right)\right)\right)}{\tau _P^4+\tau _C^4 \left(1-\tau _J \tau _P+\tau
   _F^2 \tau _P^2\right){}^2+\tau _C^2 \tau _P^2 \left(2-2 \tau _J \tau _P+\left(-2 \tau _F^2+\tau
   _J^2\right) \tau _P^2\right)}, \nonumber\\ \mathcal{D}_4&=&\frac{v_0 \tau _C^2 \tau _P \left(\tau _P^2 \left(-\tau _J+2 \tau _F^2 \tau _P\right)-\tau _C^2
   \left(-\tau _J+\left(-2 \tau _F^2+\tau _J^2\right) \tau _P\right) \left(-1+\tau _P \left(\tau _J-\tau
   _F^2 \tau _P\right)\right)\right)}{\sqrt{4 \tau _F^2-\tau _J^2} \left(\tau _P^4+\tau _C^4 \left(1-\tau
   _J \tau _P+\tau _F^2 \tau _P^2\right){}^2+\tau _C^2 \tau _P^2 \left(2-2 \tau _J \tau _P+\left(-2 \tau
   _F^2+\tau _J^2\right) \tau _P^2\right)\right)},\nonumber\\ \mathcal{E}_4&=&-\frac{v_0 \tau _C \tau _P^2 \left(\tau _P^2+\tau _C^2 \left(1-2 \tau _J \tau _P+\left(-\tau _F^2+\tau
   _J^2\right) \tau _P^2\right)\right)}{\tau _P^4+\tau _C^4 \left(1-\tau _J \tau _P+\tau _F^2 \tau
   _P^2\right){}^2+\tau _C^2 \tau _P^2 \left(2-2 \tau _J \tau _P+\left(-2 \tau _F^2+\tau _J^2\right) \tau
   _P^2\right)}
,\nonumber\\ \mathcal{F}_4&=&\frac{v_0 \tau _C \tau _P^2 \left(-\tau _J \tau _P^2-\tau _C^2 \left(4 \tau _F^2 \tau _P+\tau _J
   \left(1-2 \tau _J \tau _P+\left(-3 \tau _F^2+\tau _J^2\right) \tau _P^2\right)\right)\right)}{\sqrt{4
   \tau _F^2-\tau _J^2} \left(\tau _P^4+\tau _C^4 \left(1-\tau _J \tau _P+\tau _F^2 \tau
   _P^2\right){}^2+\tau _C^2 \tau _P^2 \left(2-2 \tau _J \tau _P+\left(-2 \tau _F^2+\tau _J^2\right) \tau
   _P^2\right)\right)}.
   \end{eqnarray}

\bibliographystyle{iopart-num}
\bibliography{main}

\end{document}